\documentclass[a4paper,11pt]{article}
\usepackage{pos}
\usepackage{float}  
\usepackage{wrapfig}
\usepackage{subcaption}
\usepackage{stackengine}

\title{HMC and gradient flow with machine-learned classically perfect fixed-point actions}
\ShortTitle{HMC and gradient flow with machine-learned classically perfect FP actions}

\author[a]{K.~Holland}
\author[b]{A.~Ipp}
\author[b]{D.~I.~M\"uller}
\author*[c]{U.~Wenger}

\affiliation[a]{University of the Pacific,
3601 Pacific Ave., Stockton, CA 95211, USA}
\affiliation[b]{Institute for Theoretical Physics, TU Wien, Wiedner Hauptstraße  8-10/136, A-1040 Vienna, Austria}
\affiliation[c]{Institute for Theoretical Physics, Albert Einstein Center for Fundamental Physics, University of Bern, CH-3012 Bern, Switzerland}

\emailAdd{wenger@itp.unibe.ch}

\abstract{Fixed-point (FP) lattice actions are classically perfect, i.e., they have continuum classical properties unaffected by discretization effects and are expected to have suppressed lattice artifacts at weak coupling. Therefore they provide a possible way to extract continuum physics with coarser lattices, allowing to circumvent problems with critical slowing down and topological freezing towards the continuum limit. We use machine-learning methods to parameterize a FP action for four-dimensional SU(3) gauge theory using lattice gauge-covariant convolutional neural networks. The large operator space allows us to find superior parameterizations compared to previous studies and we show how such actions can be efficiently simulated with the Hybrid Monte Carlo algorithm. Furthermore, we argue that FP lattice actions can be used to define a classically perfect gradient flow without any lattice artifacts at tree level. We present initial results for scaling of the gradient flow with the FP action.}

\FullConference{The 41st International Symposium on Lattice Field Theory (LATTICE2024)\\
 28 July - 3 August 2024\\
Liverpool, UK\\}

\begin{document}
\maketitle

\section{Introduction}

\begin{figure}[h!]
        \centering
        \includegraphics[width=\linewidth]{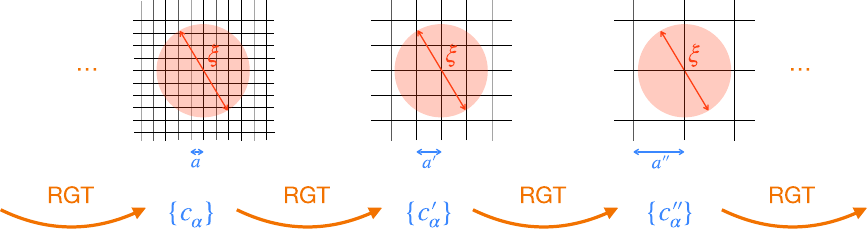}
        \caption{Illustration of the continuum limit towards the left with lattice spacings $a/\xi < a'/\xi < a''/\xi$ at fixed physical scale $\xi$ ({\it upper row}), and  renormalization group transformation (RGT) steps yielding effective couplings $\{c_\alpha\} \rightarrow \{c'_\alpha\} \rightarrow \{c''_\alpha\}$ ({\it lower row}).}
        \label{fig:RGT_illustration}
    \end{figure}

One of the most difficult challenges in lattice QCD simulations is the critical slowing down towards the continuum limit where the lattice spacing $a$ is taken to zero while keeping some physical scales fixed, cf.~Fig.~\ref{fig:RGT_illustration} for an illustration. The critical behaviour manifests itself in long autocorrelation times in the measurement of physical observables~\cite{Schaefer:2010hu}, for example gradient-flow quantities or the topological charge. The slow decorrelation in the latter quantity is usually referred to as topological freezing. While the problem is absent at sufficiently large lattice spacings, simulations at coarse resolution in general lead to large lattice artifacts and correspondingly large systematic errors due to the long extrapolation to $a\rightarrow 0$, unless one uses highly improved lattice discretizations of the underlying theory. 

One way to construct highly improved lattice actions is through renormalization group transformations (RGT). The aim of an RGT 
is to describe the physics of a fine lattice on a coarse lattice by means of adjusting the effective couplings of the discretized theory, cf.~Fig.~\ref{fig:RGT_illustration}, such that the resulting effective lattice action maintains the low-energy physics exactly. 

In these proceedings we describe the status of our project to construct an RG-improved lattice action which is classically perfect, i.e., which has no lattice discretization artifacts at tree level in the gauge coupling, and to employ it in Hybrid Monte Carlo (HMC) simulations. Previous reports of this ongoing project were given in~\cite{Holland:2024muu,Holland:2023ews}.

\section{Renormalization group transformations and fixed-point actions}
\begin{figure}[t!]
    \centering
    \includegraphics[width=0.8\textwidth]{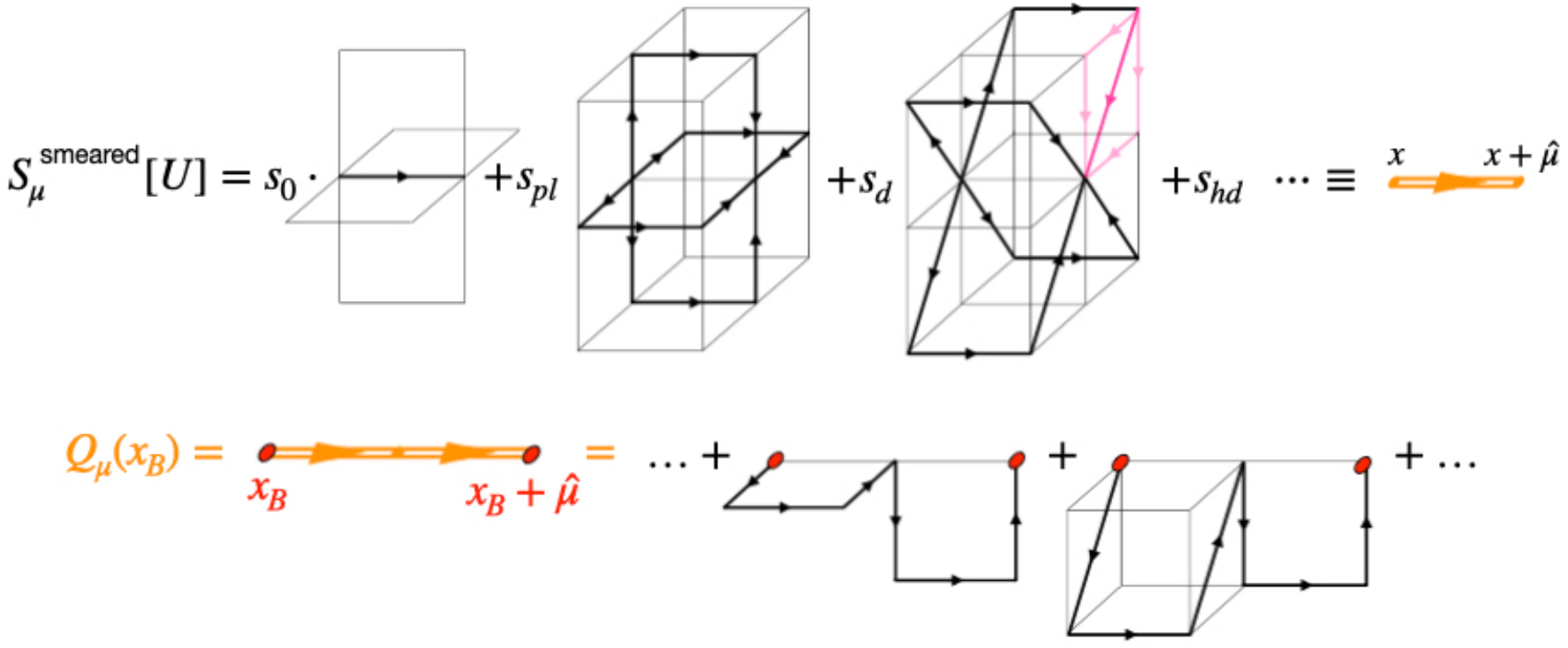}
    \caption{Illustration of the RGT blocking employed here.}
    \label{fig:RGT}
    \vspace*{-0.2cm}
\end{figure}

For gauge theories, an RGT for a lattice action $\mathcal A$ from a fine to a coarse lattice can be defined through  
\begin{equation}
    \exp(-\beta' \mathcal A'[V]) = \int \mathcal{D} U \exp( -\beta \{ \mathcal A[U] + T[U,V] \} ),
    \label{eq:rg}
\end{equation}
where the integration on the RHS~is over gauge-field configurations $U$ on the fine lattice restricted by the RG blocking kernel $T[U,V]$. In this process, the gauge coupling $\beta$ and the action $\mathcal A$ parameterized by the couplings $\{c_\alpha\}$ on the fine lattice are mapped to $\beta'$ and $\mathcal A'$, i.e., $\{c'_\alpha\}$, on the coarse lattice. The RG blocking is defined as follows. First, a smeared link $S^\text{smeared}_\mu$ is constructed  from the fine gauge links by adding up the original link $U_\mu(x)$ with weight $s_0$, the planar staples with weight $s_{pl}$, diagonal staples with weight $s_{d}$, and hyper-diagonal staples with weight $s_{hd}$, as illustrated in the top row of Fig.~\ref{fig:RGT}. Two smeared links are then multiplied together to construct the    
blocked link $Q_\mu(x_B)$ generating a combination of many gauge-link paths within 16 hypercubes of the fine lattice, cf.~bottom row of Fig.~\ref{fig:RGT}. Finally, the gauge invariant blocking kernel $T[U,V]$ is defined by
\begin{equation}
    T[U,V] =  - \kappa \sum_{n_B,\mu} \Bigl\{ {\rm Re Tr} \left(V_{n_B,\mu}Q_{n_B,\mu}^\dagger[U] \right) - \mathcal{N}^\beta_\mu[U] \Bigr\}, 
    \label{eq:rg2}
\end{equation}
where the term ${\cal N}_\mu^\beta[U]$ guarantees the correct normalization of the partition function under the RGT. The parameters $s_{0}, s_{pl}, s_{d}, s_{hd},\kappa$ of the blocking kernel can be chosen to optimize the locality of 
\begin{wrapfigure}{r}{0.7\textwidth}
    \centering
    \includegraphics[width=0.7\textwidth]{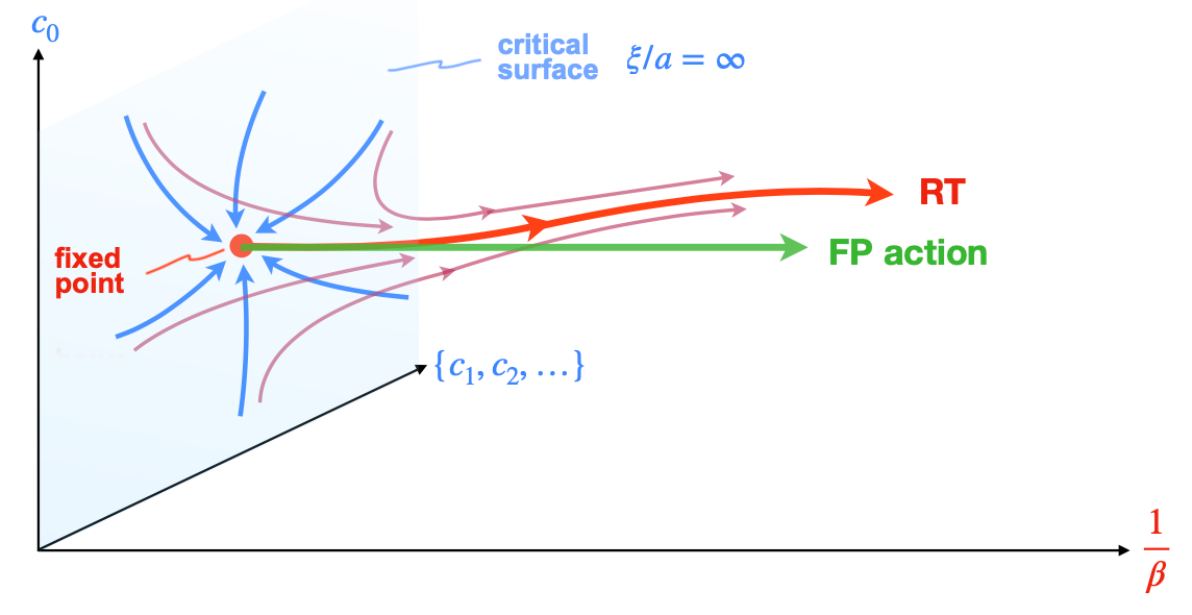}
    \caption{Illustration of the RG flow for asymptotically free gauge theories.}
    \label{fig:RGT flow}
\end{wrapfigure}
the resulting effective gauge action. For asymptotically free theories, there is one marginally relevant operator parameterized by $\beta$ and hence the critical surface, where $\xi/a = \infty$, is located at $\beta = \infty$. Under repeated RGT steps, any lattice action defined on the critical surface is driven to the Fixed Point (FP), cf.~Fig.~\ref{fig:RGT flow}.
Lattice actions slightly off the critical surface flow towards the Renormalized Trajectory (RT) which emerges from the FP along the relevant direction. Actions on the RT are effective actions with no lattice artifacts at all. 

The FP at $\beta \rightarrow \infty$ is defined by the FP equation~\cite{Hasenfratz:1993sp}
\begin{equation}
   {\mathcal A}^{\rm FP}[V] = \min_{U} \left( {\mathcal A}^{\rm FP}[U] + T[U,V] \right) \, ,
\label{eq:FP equation} 
\end{equation}
and using ${\mathcal A}^{\rm FP}$ with fixed couplings $\{c_\alpha^\text{FP}\}$ at finite $\beta$ defines the straight line in Fig.~\ref{fig:RGT flow}. It is expected to stay close to the RT at weak coupling and hence to have suppressed lattice artifacts.
While the FP Eq.~(\ref{eq:FP equation}) implicitly provides data for each coarse configuration $V$, finding the explicit form of the FP action, i.e., the couplings $\{c_\alpha^\text{FP}\}$, is a difficult parameterization problem. It can be addressed efficiently using convolutional neural networks and machine-learning techniques.

\section{Neural network parameterization and HMC simulation}
\begin{figure}[t!]
\centering
 \includegraphics[width=0.75\textwidth]{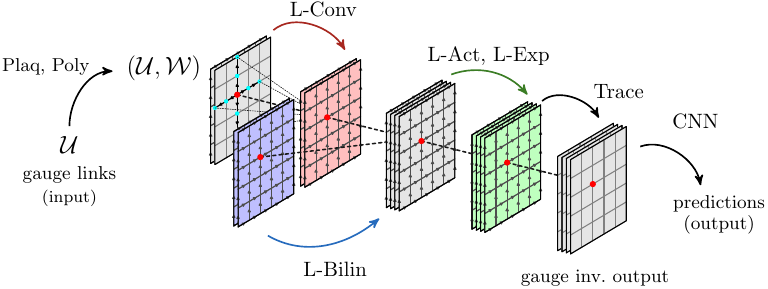}
  \caption{Example of an L-CNN that processes lattice gauge data while preserving gauge symmetry through specialized convolutional (L-Conv), bilinear (L-Bilin), and activation layers~\cite{Favoni:2020reg}. }
  \label{fig:architecture}
\end{figure}

Since the FP action in general involves infinitely many gauge invariant operators with their corresponding couplings $\{c_\alpha^\text{FP}\}$, any useful parameterization requires high flexibility while maintaining exact gauge invariance. The lattice gauge-covariant convolutional neural network (L-CNN) architecture described in Ref.~\cite{Favoni:2020reg} is well 
suited for this task. Taking as input a gauge field $\cal U$ and gauge covariant objects $\cal W$, e.g., untraced Wilson
plaquettes and Polyakov loops, the network produces arbitrarily complicated loops through convolutional layers with parallel transports, while additional bilinear layers combine products of covariant objects locally, retaining the gauge symmetry at each step. Finally, a trace layer produces gauge invariant output, cf.~Fig.~\ref{fig:architecture} for an illustration. 
We employ machine-learning techniques to train the network to reproduce the FP action values given by the RHS of Eq.~(\ref{eq:FP equation}) by parameterizing the FP action on the LHS. 
In addition, the derivatives of the FP action with respect to the gauge links are also determined by the FP equation, and they provide $8 \times 4 \times (\mathrm{volume})$ data per gauge configuration. The L-CNN generates these derivatives exactly through backpropagation, hence these large datasets are essential in training the network.

\begin{figure}[ht]
\footnotesize   
    \stackunder[5pt]{\includegraphics[width=0.48\textwidth]{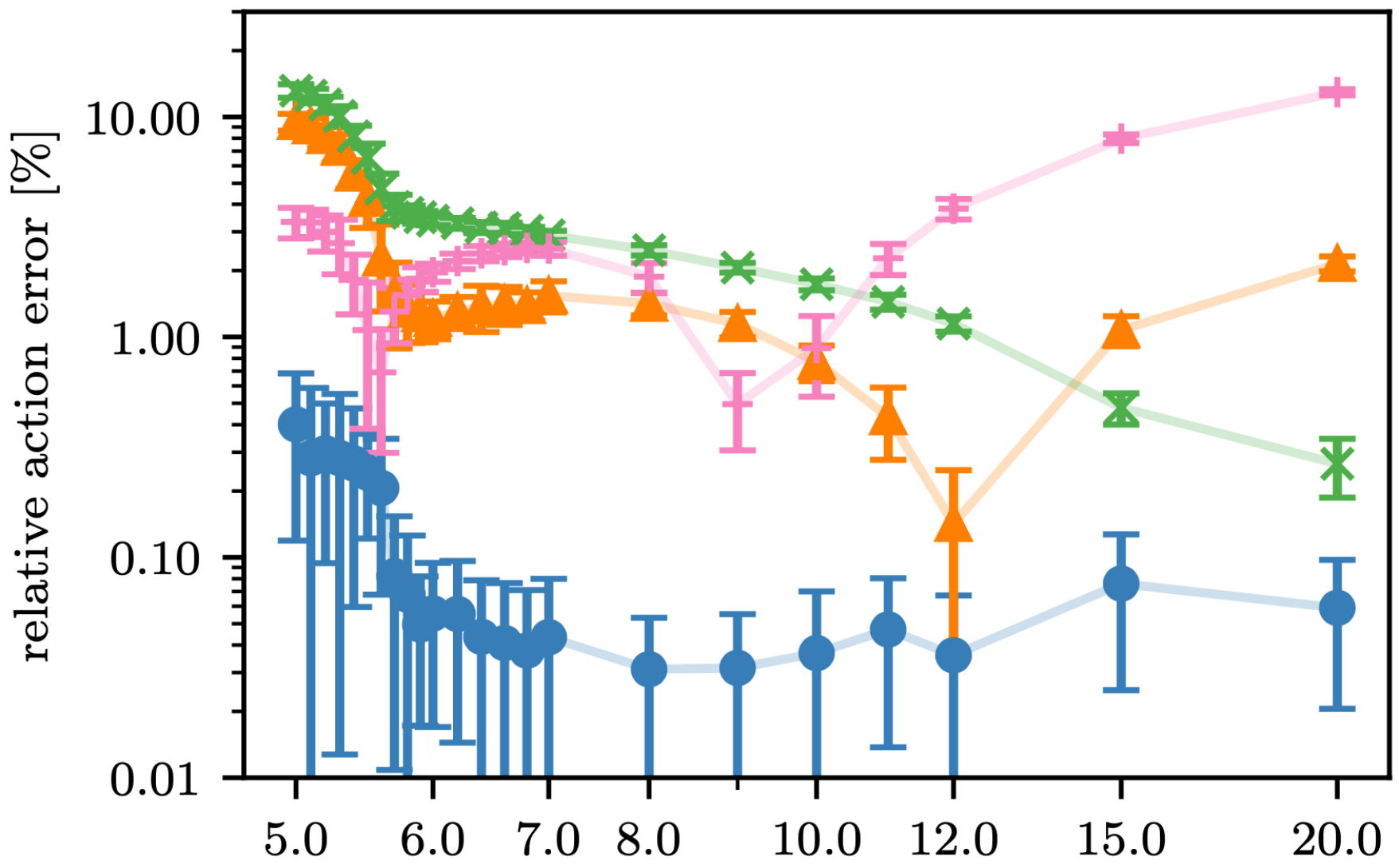}}{\hspace*{1.2cm}$\beta_\text{wil}$}
    \stackunder[5pt]{\includegraphics[width=0.48\textwidth]{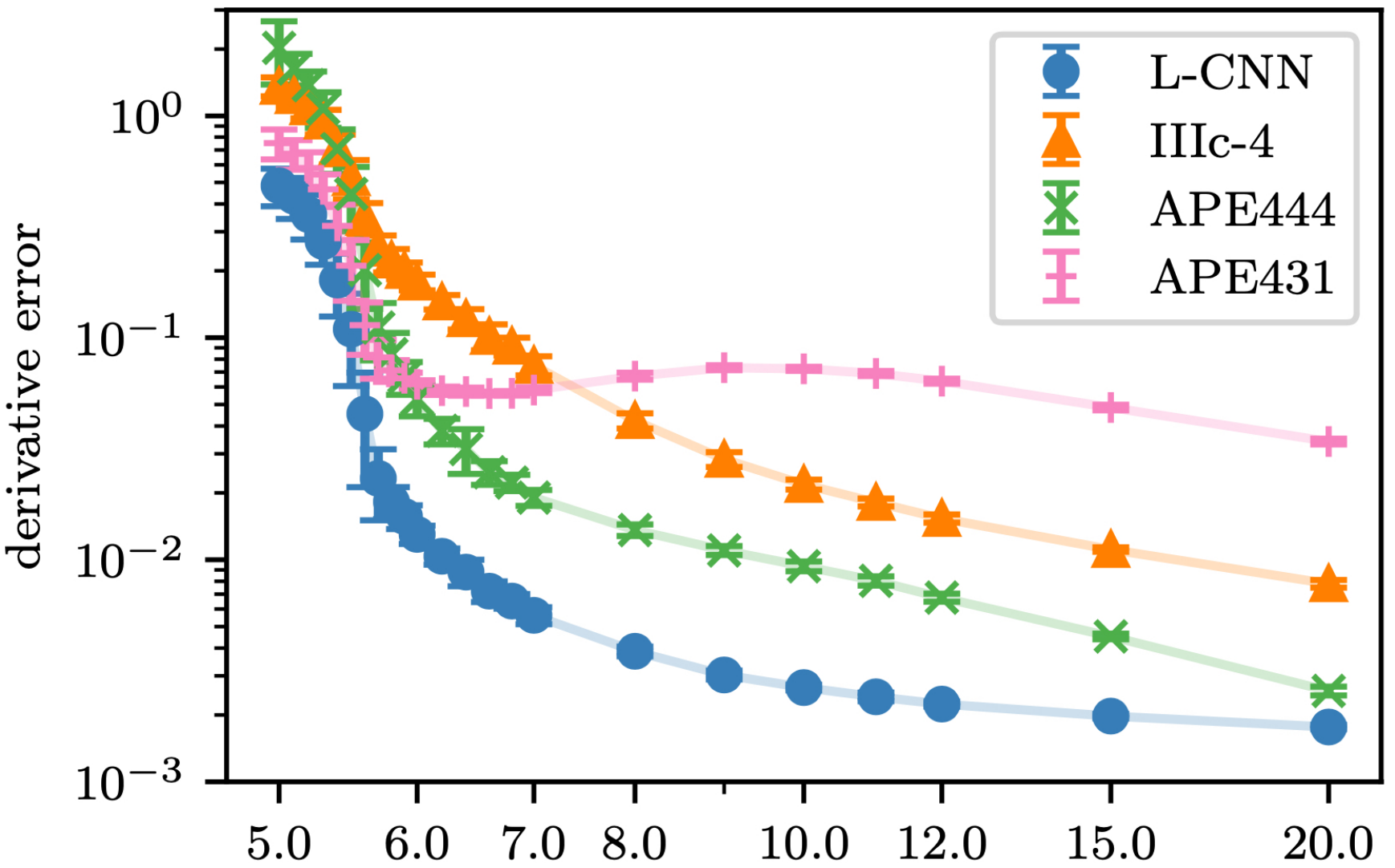}}{\hspace*{1.2cm}$\beta_\text{wil}$}
    \caption{Comparison of the new parameterization of the FP action (L-CNN) and previous parameterizations using either powers of traced loops up to length six (IIIc-4) or powers of traced plaquettes containing APE-smeared fat links (APE 431, APE 444), from Ref.~\cite{Holland:2024muu}.}
    \label{fig:best model}
\end{figure}
As our best parameterization, 
we use an L-CNN consisting of three layers with 12, 24, 24 channels and kernel sizes 2, 2, 1. In Fig.~\ref{fig:best model} we show the comparison with previous 
parameterizations based on traced loops up to length six and up to four powers of the traces (IIIc-4) \cite{Blatter:1996np}, and parameterizations based on powers of traced plaquettes constructed from hypercubically APE-smeared fat links (APE444, APE431) \cite{Niedermayer:2000yx}. 
To test the quality of different parameterizations, we take as input ensembles of coarse configurations generated with the Wilson gauge action at a range of $\beta_{\mathrm{wil}}$ values to cover a broad span of lattice spacings, where numerical minimization on the RHS of Eq.~(\ref{eq:FP equation}) provides the underlying true action values. The left panel shows the relative error of the parameterized FP action values on these ensembles at each $\beta_{\mathrm{wil}}$
indicated on the $x$-axis, while the right panel shows the size of the parameterization errors of the derivatives. We find that the L-CNN consistently performs better than the previous parameterizations all the way from coarse to fine lattice spacing. 

While an accurate parameterization is a necessary first step in the FP approach, there is no guarantee that the lattice artifacts will in fact be small when this action is used for nonperturbative calculations. This can only be tested with an actual scaling study involving MC simulations 
\begin{wrapfigure}{l}{0.5\textwidth}
    \includegraphics[width=.5\textwidth]{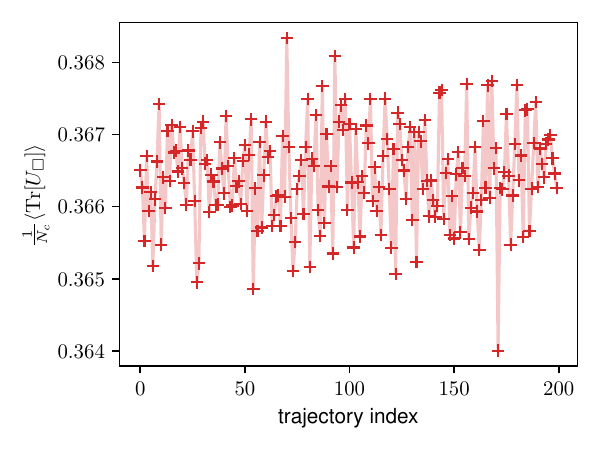}
    \caption{HMC evolution of the plaquette value for the FP action at $\beta_\text{FP}=2.95$ corresponding to $a \simeq 0.11$~fm.}
    \label{fig:HMC}
\end{wrapfigure}
 at a range of $\beta$ values, to make quantitative comparison with other lattice actions and the continuum limit of appropriately chosen quantities. Given that the derivatives of the FP action with respect to gauge links are automatically provided by the L-CNN through backpropagation, it is natural to choose the Hybrid Monte Carlo (HMC) algorithm~\cite{Duane:1987de,Gottlieb:1987mq} for simulations with the FP action, adding fictitious conjugate momenta to evolve the gauge links according to the equations of motion. Standard techniques for accelerating the HMC algorithm can be used for the FP action, for example we use the 4MN4FP variant~\cite{Takaishi:2005tz} of the 4th order Omelyan integrator~\cite{Omelyan:2003bjg} to allow for a coarser time step when solving the equations of motion. Among the tests of our implementation, we check that the Hamiltonian constraint $\langle \exp(-\Delta H) \rangle = 1$ is satisfied by ensembles generated with the HMC algorithm.
We show in Fig.~\ref{fig:HMC} an example of the HMC evolution of the plaquette values for a stream of 200 trajectories obtained by simulating the FP action at $\beta_\text{FP}= 2.95$, which corresponds to a lattice spacing of $a \simeq 0.11$~fm.

\section{Classically perfect gradient flow}

A precise observable is necessary to test the quality of improvement of the FP action, without introducing additional lattice artifacts. This is naturally given by the gradient flow (GF) of the gauge fields~\cite{Luscher:2010iy,Luscher:2011bx}. In continuum language, given the gauge action $\mathcal{A} = (1/g^2) \int d^4x (- \frac{1}{2} {\rm Tr} F_{\mu \nu}F_{\mu \nu})$, with the field strength $F_{\mu \nu} = \partial_\mu A_\nu - \partial_\nu A_\mu + [A_\mu,A_\nu]$ and gauge coupling $g$, the non-Abelian gauge fields $A_\mu(t)$ evolve with a fictitious time $t$ as 
\begin{equation}
    \frac{d A_\mu(t)}{dt} = - \frac{\delta {\cal A}}{\delta A_\mu(t)},
\end{equation}
where the flow starts with the original gauge field $A_\mu(t=0)$ and the flow time $t$ has dimension ({{\it{length}})$^2$. The resulting action density of flowed gauge fields $E(t)=- \frac{1}{2} {\rm Tr} F_{\mu \nu}F_{\mu \nu}$ is  renormalized, which allows a number of connections to physical quantities. For example, perturbatively the flowed action density for SU($N$) gauge theory is 
\begin{equation}
    t^2 \langle E(t) \rangle = \frac{3(N^2-1)g^2}{128 \pi^2}( 1 + {\cal O}(g^2) ), 
\end{equation}
where $g$ is the renormalized gauge coupling in the $\overline{\rm MS}$ scheme with RG scale $\mu = 1/\sqrt{8t}$. Hence a specific value of $g^2(t)$ implicitly defines a value for the flow time $t$ in physical units, which allows the scale of lattice simulations to be determined. Another use is the gauge coupling $\beta$-function defined through the GF scheme, e.g., for SU(3) gauge theory
\begin{equation}
    t^2 \langle E(t) \rangle \doteq \frac{3 g_{\rm GF}^2(t)}{16 \pi^2},
\end{equation}
where the $\beta$-function $\mu^2 \cdot dg^2_{\rm GF}/d(\mu^2) = -t \cdot dg^2_{\rm GF}/dt$ can be measured directly from the GF on  gauge configurations generated at chosen values of the coupling~\cite{Fodor:2017die,Hasenfratz:2019hpg}. In general, GF quantities can be measured with high accuracy, making them good probes of discretization effects in lattice simulations.  

On the lattice, the flow is of the gauge links, with $dU_\mu/dt = - i (\delta {\cal A}^f/\delta U_\mu) U_\mu$ for some choice ${\cal A}^f$ of lattice flow action. In addition, one has a separate choice of lattice action ${\cal A}^e$ for the discretized action density observable $E(t)$. The gauge configurations are generated via MC simulation, where there is a third choice of discretized action ${\cal A}^g$. The collective discretization artifacts of these choices can be calculated perturbatively~\cite{Fodor:2014cpa}, with
\begin{eqnarray}
t^2 \langle E(t) \rangle &=& \frac{3(N^2-1)g_0^2}{128 \pi^2}\left[ C(a^2/t) + {\cal O}(g_0^2) \right], \nonumber \\
        C(a^2/t) 
        &=& \frac{64 \pi^2 t^2}{3} 
        \int_{-\pi/a}^{\pi/a} \frac{d^4 p}{(2 \pi)^4} {\rm Tr}\left[e^{-t({\cal A}^f+{\cal G})}({\cal A}^g+{\cal G})^{-1}e^{-t({\cal A}^f+{\cal G})} {\cal A}^e \right],
\end{eqnarray}
where $C(a^2/t)$ contains the tree-level lattice artifacts as deviations from 1,
the action density in momentum space to quadratic order in the fields is 
$A_{\mu}(p) {\cal A}_{\mu \nu}(p) A_\nu(-p)$, $g_0$ is the bare gauge coupling and ${\cal G}$ is a gauge fixing term. One can exploit the freedom to choose different gauge actions for ${\cal A}^e, {\cal A}^f$ and ${\cal A}^g$ to minimize the artifacts in a tree-level improved renormalized gauge coupling~\cite{Fodor:2014cpa}. On the other hand, using the same lattice action for all three gives
\begin{equation}
    C(a^2/t) = \frac{64 \pi^2 t^2}{3} \int_{-\pi/a}^{\pi/a} \frac{d^4 p}{(2 \pi)^4} {\rm Tr}[e^{-2t({\cal A}+{\cal G})}].
    \label{eq:coeff}
\end{equation}

    \begin{figure}
        \includegraphics[width=0.48\textwidth]{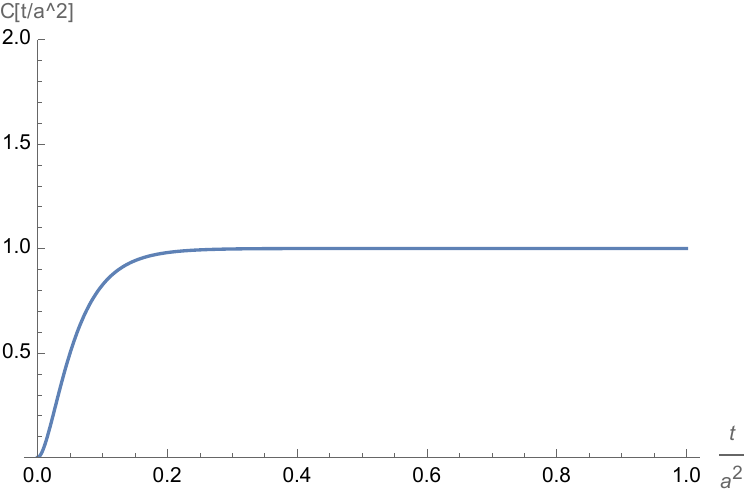}
        \includegraphics[width=0.48\textwidth]{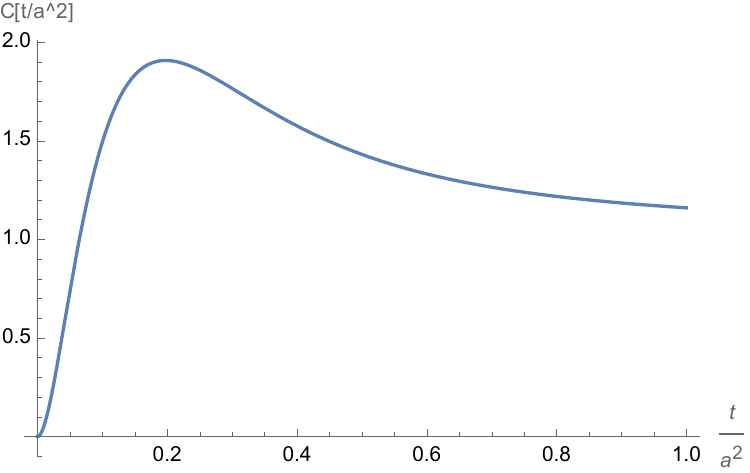} 
        \caption{The tree-level coefficient $C(a^2/t)$ in the gauge field gradient flow for different regulators, with a hard momentum cutoff ({\it left}) or  Wilson lattice action ({\it right}). The exact FP propagator extends the momentum range to $\pm \infty$ even at finite lattice spacing, resulting in $C^{\mathrm {FP}}(a^2/t)=1$ with no artifacts.}
        \label{fig:GF_corrections}
    \end{figure}

As a pedagogical guide, examples of the tree-level coefficient $C(a^2/t)$ for two different regulators are shown in Fig.~\ref{fig:GF_corrections}. One is a hard-momentum cutoff $\pm \pi/a$ with the continuum dispersion relation, the second is the Wilson lattice gauge action for which $C(a^2/t)$ is analytically calculable in terms of the Bessel function $I_0$. For both, the coefficient approaches 1 in the limit $t/a^2 \rightarrow \infty$. 

For the exact FP action, the FP propagator is substituted into Eq.~(\ref{eq:coeff}). At any finite lattice spacing, the gauge configuration can be connected back to an arbitrarily fine initial configuration through multiple RG transformations, where the propagator is also RG blocked. Although on the lattice the momenta are restricted to $-\pi/a \le p_\mu \le \pi/a$, the iterated RG transformations result in the FP propagator having poles at $( p + 2 \pi l)^2$ for all integers $l$. Hence the momentum range is extended to $\pm \infty$ and the exact FP propagator has the continuum dispersion relation. This results in $C^{\mathrm{FP}}(a^2/t) = 1$, i.e., the exact FP action has no tree-level artifacts in the gradient flow. This new finding is in keeping with FP actions preserving classical properties exactly on the lattice, e.g., the continuum dispersion relation, exact classical (instanton) solutions and invariant topological charge under RG transformations, and chirality of FP fermions as a solution of the Ginsparg-Wilson relation~\cite{DeGrand:1995ji,Hasenfratz:1998jp,Hasenfratz:1998ri}.

\section{Scaling of gradient-flow scales}
Having generated MC ensembles with the FP action, we integrate the flow equation on each configuration with the same FP action using a third-order Runge-Kutta scheme~\cite{Luscher:2010iy}. 
In the left plot of Fig.~\ref{fig:scaling} we show the resulting flow of the action density for one FP gauge ensemble as an example.  Physical reference scales are defined through the conditions~\cite{Luscher:2010iy,BMW:2012hcm} 
\begin{equation}
t^2 \langle E \rangle |_{t={t_{0.3}}}=0.3, \qquad t \frac{d}{dt} \left( t^2 \langle E \rangle \right) \bigg|_{t=w_{0.3}^2} = 0.3 \,,
\end{equation}
and we use the scaling of a dimensionless ratio of these scales to probe the FP action for any remaining lattice artifacts. With the classically perfect GF at hand, any such lattice artifacts are due either to an imperfect parameterization or to quantum effects. In the right plot of Fig.~\ref{fig:scaling} we show the scaling of the ratio $t_{0.3}/w_{0.3}^2$ towards the continuum limit $a^2/t_{0.3} \rightarrow 0$ for the parameterized FP action and compare it with that of the Wilson gauge action using either the plaquette or clover discretization of the observable $E(t)$. We use $w_{0.3}  = 0.1755$~fm~\cite{BMW:2012hcm} for a visual guide of the lattice spacing. Discretization effects in the parameterized FP action are very mild and an ${\cal O}(a^2)$ extrapolation of the data with $a \lesssim 0.13$~fm is sufficient for a reliable continuum limit. This is in contrast to the Wilson gauge action which requires simulations at substantially smaller lattice spacings for an equally reliable extrapolation.

\begin{figure}
    \centering
    \includegraphics[width=0.46\linewidth]{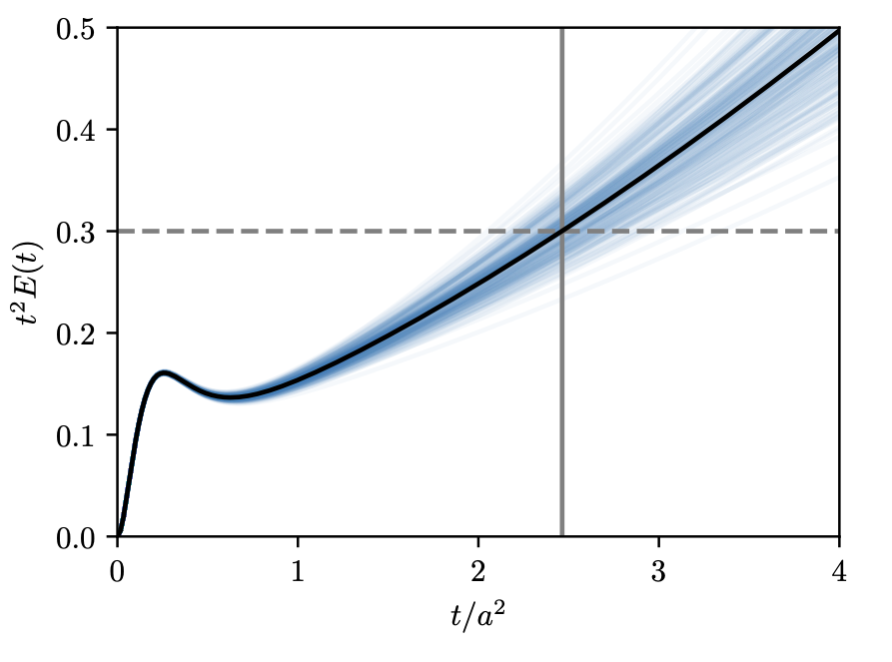}
\includegraphics[width=0.475\textwidth]{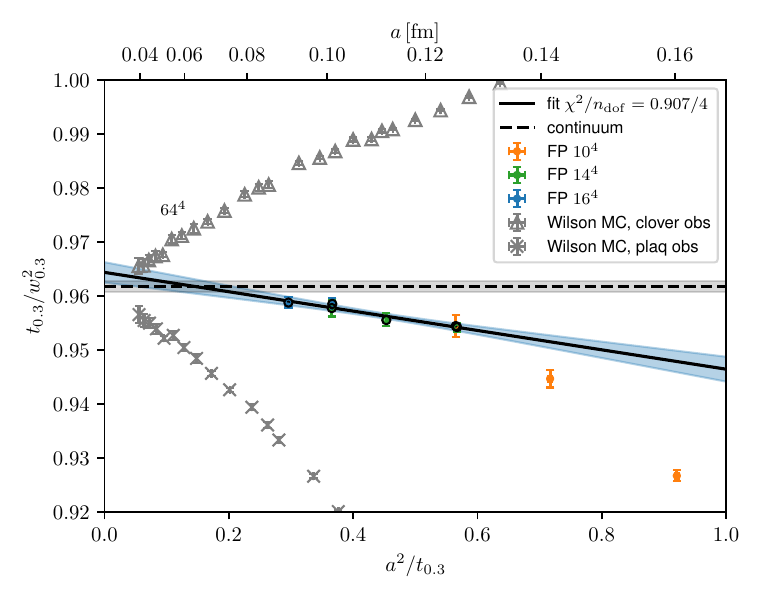}
    \caption{({\it Left}) Example of gradient flow on one FP gauge ensemble, the light blue curves are from individual configurations, the black curve is the ensemble average. 
    ({\it Right}) Scaling test of FP simulation results for the combination of scales $t_{0.3}/w_{0.3}^2$. 
    }
    \label{fig:scaling}
\end{figure}

\section{Conclusions and outlook}

In this work, we have demonstrated how a fixed-point lattice gauge action can be parameterized using modern machine-learning tools, argued that the exact FP gradient flow is free of tree-level lattice artifacts, and presented preliminary results for a scaling test of the parameterized FP action through MC simulations and measurements of the gradient flow in four-dimensional SU(3) gauge theory. Lattice gauge-covariant convolutional neural networks (L-CNNs) are an essential element to achieve a superior parameterization of the FP action with exact gauge invariance. The initial results show that lattice artifacts in the parameterized FP action are much reduced, making it broadly useful for any gauge theory study, allowing continuum properties to be extracted from simulations on coarser lattices.

\acknowledgments

This work is supported in part by the Swiss National Science Foundation (SNSF) through grant No.200020$\_$208222. 
K.H.~wishes to thank the AEC and ITP at the University of Bern for their support. 
A.I.~and D.M.~acknowledge
funding from the Austrian Science Fund (FWF) projects
P 32446, P 34455 and P 34764.
K.H.~acknowledges funding from the US National Science Foundation under grants 2014150 and 2412320. 
The computational results presented have been achieved in part using the Vienna Scientific Cluster (VSC) and LEONARDO at CINECA, Italy, via an AURELEO (Austrian Users at LEONARDO supercomputer) project, and UBELIX (\url{http://www.id.unibe.ch/hpc}), the HPC cluster at the University of Bern.

\bibliographystyle{JHEP}
\bibliography{references}

\providecommand{\href}[2]{#2}\begingroup\raggedright\begin{thebibliography}{10}

\bibitem{Schaefer:2010hu}
{\scshape ALPHA} collaboration, S.~Schaefer, R.~Sommer and F.~Virotta,
  \emph{{Critical slowing down and error analysis in lattice QCD simulations}},
  \href{http://dx.doi.org/10.1016/j.nuclphysb.2010.11.020}{\emph{Nucl. Phys. B}
  {\bf 845} (2011) 93}, [\href{http://arxiv.org/abs/1009.5228}{{\tt
  1009.5228}}].

\bibitem{Holland:2024muu}
K.~Holland, A.~Ipp, D.~I. M\"uller and U.~Wenger, \emph{{Machine learning a
  fixed point action for SU(3) gauge theory with a gauge equivariant
  convolutional neural network}},
  \href{http://dx.doi.org/10.1103/PhysRevD.110.074502}{\emph{Phys. Rev. D} {\bf
  110} (2024) 074502}, [\href{http://arxiv.org/abs/2401.06481}{{\tt
  2401.06481}}].

\bibitem{Holland:2023ews}
K.~Holland, A.~Ipp, D.~I. M\"uller and U.~Wenger, \emph{{Fixed point actions
  from convolutional neural networks}},
  \href{http://dx.doi.org/10.22323/1.453.0038}{\emph{PoS} {\bf LATTICE2023}
  (2024) 038}, [\href{http://arxiv.org/abs/2311.17816}{{\tt 2311.17816}}].

\bibitem{Hasenfratz:1993sp}
P.~Hasenfratz and F.~Niedermayer, \emph{{Perfect lattice action for
  asymptotically free theories}},
  \href{http://dx.doi.org/10.1016/0550-3213(94)90261-5}{\emph{Nucl. Phys. B}
  {\bf 414} (1994) 785}, [\href{http://arxiv.org/abs/hep-lat/9308004}{{\tt
  hep-lat/9308004}}].

\bibitem{Favoni:2020reg}
M.~Favoni, A.~Ipp, D.~I. M\"uller and D.~Schuh, \emph{{Lattice Gauge
  Equivariant Convolutional Neural Networks}},
  \href{http://dx.doi.org/10.1103/PhysRevLett.128.032003}{\emph{Phys. Rev.
  Lett.} {\bf 128} (2022) 032003}, [\href{http://arxiv.org/abs/2012.12901}{{\tt
  2012.12901}}].

\bibitem{Blatter:1996np}
M.~Blatter and F.~Niedermayer, \emph{{New fixed point action for SU(3) lattice
  gauge theory}},
  \href{http://dx.doi.org/10.1016/S0550-3213(96)00523-8}{\emph{Nucl. Phys. B}
  {\bf 482} (1996) 286}, [\href{http://arxiv.org/abs/hep-lat/9605017}{{\tt
  hep-lat/9605017}}].

\bibitem{Niedermayer:2000yx}
F.~Niedermayer, P.~Rufenacht and U.~Wenger, \emph{{Fixed point gauge actions
  with fat links: Scaling and glueballs}},
  \href{http://dx.doi.org/10.1016/S0550-3213(00)00731-8}{\emph{Nucl. Phys. B}
  {\bf 597} (2001) 413}, [\href{http://arxiv.org/abs/hep-lat/0007007}{{\tt
  hep-lat/0007007}}].

\bibitem{Duane:1987de}
S.~Duane, A.~D. Kennedy, B.~J. Pendleton and D.~Roweth, \emph{{Hybrid Monte
  Carlo}}, \href{http://dx.doi.org/10.1016/0370-2693(87)91197-X}{\emph{Phys.
  Lett. B} {\bf 195} (1987) 216}.

\bibitem{Gottlieb:1987mq}
S.~A. Gottlieb, W.~Liu, D.~Toussaint, R.~L. Renken and R.~L. Sugar,
  \emph{{Hybrid Molecular Dynamics Algorithms for the Numerical Simulation of
  Quantum Chromodynamics}},
  \href{http://dx.doi.org/10.1103/PhysRevD.35.2531}{\emph{Phys. Rev. D} {\bf
  35} (1987) 2531}.

\bibitem{Takaishi:2005tz}
T.~Takaishi and P.~de~Forcrand, \emph{{Testing and tuning new symplectic
  integrators for hybrid Monte Carlo algorithm in lattice QCD}},
  \href{http://dx.doi.org/10.1103/PhysRevE.73.036706}{\emph{Phys. Rev. E} {\bf
  73} (2006) 036706}, [\href{http://arxiv.org/abs/hep-lat/0505020}{{\tt
  hep-lat/0505020}}].

\bibitem{Omelyan:2003bjg}
I.~P. Omelyan, I.~M. Mryglod and R.~Folk, \emph{{Symplectic analytically
  integrable decomposition algorithms: classification, derivation, and
  application to molecular dynamics, quantum and celestial mechanics
  simulations}},
  \href{http://dx.doi.org/10.1016/s0010-4655(02)00754-3}{\emph{Comput. Phys.
  Commun.} {\bf 151} (2003) 272}.

\bibitem{Luscher:2010iy}
M.~L\"uscher, \emph{{Properties and uses of the Wilson flow in lattice QCD}},
  \href{http://dx.doi.org/10.1007/JHEP08(2010)071}{\emph{JHEP} {\bf 08} (2010)
  071}, [\href{http://arxiv.org/abs/1006.4518}{{\tt 1006.4518}}].

\bibitem{Luscher:2011bx}
M.~Luscher and P.~Weisz, \emph{{Perturbative analysis of the gradient flow in
  non-abelian gauge theories}},
  \href{http://dx.doi.org/10.1007/JHEP02(2011)051}{\emph{JHEP} {\bf 02} (2011)
  051}, [\href{http://arxiv.org/abs/1101.0963}{{\tt 1101.0963}}].

\bibitem{Fodor:2017die}
Z.~Fodor, K.~Holland, J.~Kuti, D.~Nogradi and C.~H. Wong, \emph{{A new method
  for the beta function in the chiral symmetry broken phase}},
  \href{http://dx.doi.org/10.1051/epjconf/201817508027}{\emph{EPJ Web Conf.}
  {\bf 175} (2018) 08027}, [\href{http://arxiv.org/abs/1711.04833}{{\tt
  1711.04833}}].

\bibitem{Hasenfratz:2019hpg}
A.~Hasenfratz and O.~Witzel, \emph{{Continuous renormalization group $\beta$
  function from lattice simulations}},
  \href{http://dx.doi.org/10.1103/PhysRevD.101.034514}{\emph{Phys. Rev. D} {\bf
  101} (2020) 034514}, [\href{http://arxiv.org/abs/1910.06408}{{\tt
  1910.06408}}].

\bibitem{Fodor:2014cpa}
Z.~Fodor, K.~Holland, J.~Kuti, S.~Mondal, D.~Nogradi and C.~H. Wong, \emph{{The
  lattice gradient flow at tree-level and its improvement}},
  \href{http://dx.doi.org/10.1007/JHEP09(2014)018}{\emph{JHEP} {\bf 09} (2014)
  018}, [\href{http://arxiv.org/abs/1406.0827}{{\tt 1406.0827}}].

\bibitem{DeGrand:1995ji}
T.~A. DeGrand, A.~Hasenfratz, P.~Hasenfratz and F.~Niedermayer, \emph{{The
  Classically perfect fixed point action for SU(3) gauge theory}},
  \href{http://dx.doi.org/10.1016/0550-3213(95)00458-5}{\emph{Nucl. Phys. B}
  {\bf 454} (1995) 587}, [\href{http://arxiv.org/abs/hep-lat/9506030}{{\tt
  hep-lat/9506030}}].

\bibitem{Hasenfratz:1998jp}
P.~Hasenfratz, \emph{{Lattice QCD without tuning, mixing and current
  renormalization}},
  \href{http://dx.doi.org/10.1016/S0550-3213(98)00399-X}{\emph{Nucl. Phys. B}
  {\bf 525} (1998) 401}, [\href{http://arxiv.org/abs/hep-lat/9802007}{{\tt
  hep-lat/9802007}}].

\bibitem{Hasenfratz:1998ri}
P.~Hasenfratz, V.~Laliena and F.~Niedermayer, \emph{{The Index theorem in QCD
  with a finite cutoff}},
  \href{http://dx.doi.org/10.1016/S0370-2693(98)00315-3}{\emph{Phys. Lett. B}
  {\bf 427} (1998) 125}, [\href{http://arxiv.org/abs/hep-lat/9801021}{{\tt
  hep-lat/9801021}}].

\bibitem{BMW:2012hcm}
{\scshape BMW} collaboration, S.~Bors\'anyi, S.~D\"urr, Z.~Fodor, C.~Hoelbling,
  S.~D. Katz, S.~Krieg et~al., \emph{{High-precision scale setting in lattice
  QCD}}, \href{http://dx.doi.org/10.1007/JHEP09(2012)010}{\emph{JHEP} {\bf 09}
  (2012) 010}, [\href{http://arxiv.org/abs/1203.4469}{{\tt 1203.4469}}].

\end{thebibliography}\endgroup

\end{document}